\documentclass[twocolumn,prl,floatfix,superscriptaddress,nobibnotes]{revtex4-1}

\usepackage{amsmath}
\usepackage{amssymb}
\usepackage{graphicx}
\usepackage{bm}
\usepackage{color}
\usepackage[separate-uncertainty=true,alsoload=torr]{siunitx}

\begin{document}

\title{Effective {\emph g}-factor of Subgap States in Hybrid Nanowires}

\author{S.~Vaitiek\.{e}nas}
\affiliation{Center for Quantum Devices and Station Q Copenhagen, Niels Bohr Institute, University of Copenhagen, Copenhagen, Denmark}
\author{M.-T.~Deng}
\affiliation{Center for Quantum Devices and Station Q Copenhagen, Niels Bohr Institute, University of Copenhagen, Copenhagen, Denmark}
\author{J.~Nyg\aa rd}
\affiliation{Center for Quantum Devices and Station Q Copenhagen, Niels Bohr Institute, University of Copenhagen, Copenhagen, Denmark}
\author{P.~Krogstrup}
\affiliation{Center for Quantum Devices and Station Q Copenhagen, Niels Bohr Institute, University of Copenhagen, Copenhagen, Denmark}
\author{C.~M.~Marcus}
\affiliation{Center for Quantum Devices and Station Q Copenhagen, Niels Bohr Institute, University of Copenhagen, Copenhagen, Denmark}

\date{\today}

\begin{abstract}
We use the effective $g$-factor of Andreev subgap states in an axial magnetic field to investigate how the superconducting density of states is distributed between the semiconductor core and the superconducting shell in hybrid nanowires.  We find a step-like reduction of the Andreev $g$-factor and improved hard gap with reduced carrier density in the nanowire, controlled by gate voltage. These observations are relevant for Majorana devices, which require tunable carrier density and a $g$-factor exceeding that of the parent superconductor.

\end{abstract}

\maketitle

The electronic properties of a semiconductor nanowire can be altered dramatically by contacting it to a superconductor. If the nanowire has strong spin-orbit coupling, the application of a magnetic field can induce a transition from trivial to topological superconductivity, with Majorana zero modes localized at the ends of the nanowire \cite{Lutchyn2010,Oreg2010}. The Majorana bound states (MBSs) are predicted to exhibit non-Abelian statistics, and can serve as a basis for topological quantum computing \cite{Kitaev2003,Nayak2008,Alicea2011,Aasen2016}. Following concrete theoretical proposals to generate MBSs in these systems, several experiments have reported zero-bias conductance peaks \cite{Mourik2012,Das2012,Gul2018} consistent with theoretical expectation in a number of ways.  More recently, the development of epitaxial hybrid nanowires \cite{Krogstrup2015} has improved the superconducting gap \cite{Chang2015}, making evident the coalescence of Andreev bound states (ABSs) to form the zero-bias conductance peak \cite{Deng2016, Zhang2018}. 

The rate of linear decrease of the subgap ABSs towards zero energy as a function of magnetic field defines an effective $g$-factor, denoted $g^{*}$. Inducing the topological phase using an applied field requires $g^{*}$ to exceed the $g$-factor of the proximitizing s-wave superconductor, otherwise the field will drive the whole system normal. Studies on hybrid InAs/Al nanowires found $\lvert g^*\rvert$ ranging from 4 to 50 \cite{Das2012,Deng2016,Albrecht2016}, substantially different from the bulk value, $g_{\rm InAs}\sim-15$ \cite{Pidgeon1967,Konopka1967}. Gate dependence measurements of $g^*$ have been reported in an InAs/InP core/shell quantum dot coupled to a superconductor \cite{Lee2014}, where repulsion effect from superconducting continuum suppressed $g^*$ of the spin-down branch, while $g^*$ of the spin-up branch remained around $-6$. The effective $g$-factor of a quantum dot electronic states has also been studied in non-proximitized, bare InAs nanowires. A $g$-factor fluctuating between $-2$ and $-18$ has been observed in a single-dot geometry \cite{Csonka2008}. Electric and magnetic field tunable $g$-factor has been demonstrated in a double-dot geometry \cite{Schroer2011}. Some suppression of $g^*$ can be attributed to spatial confinement \cite{Lommer1985,Kiselev1998} as shown experimentally in Ref.~\cite{Bjoerk2005}, while enhancement of $g^*$ can result from a combination of Zeeman and orbital contributions in higher subbands \cite{Winkler2017}.

In this Letter, we show that the effective $g$-factor of ABSs depends sensitively on the carrier density in the wire, controlled by electrostatic gate voltages. We interpret this observation as revealing how the superconducting density of states is distributed throughout the cross section of the hybrid system. The semiconducting InAs nanowire has large spin-orbit coupling and large negative $g$-factor, whereas the superconducting Al shell, which induces the proximity effect, has small spin-orbit coupling, and $g_{\rm Al}\sim2$. At high carrier density in the wire, subgap states predominantly reside in the nanowire, reflecting the properties of the semiconductor; as carriers in the nanowire are depleted, the remaining portion of the states are confined against the InAs/Al interface, with relatively small $g^{*}$ and strong proximity effect.

\begin{figure*}[t!]
\includegraphics[width=\linewidth]{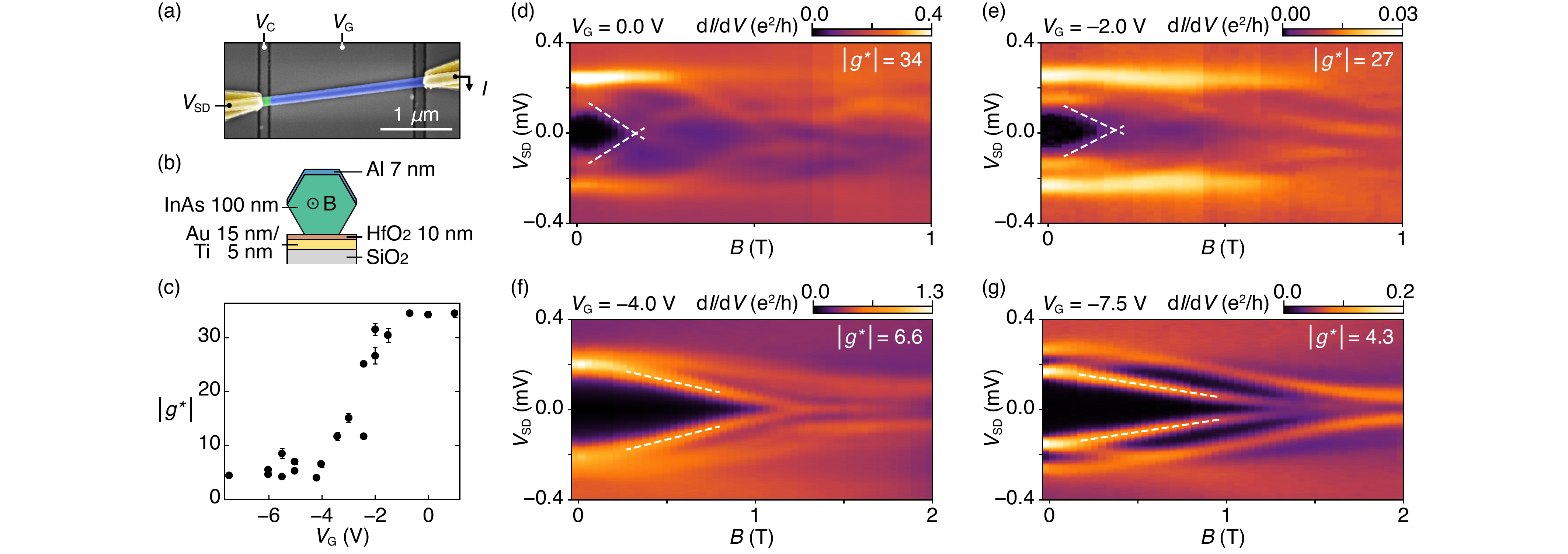}
\caption{\label{fig:1} (a) False-color electron micrograph of Device 1, showing InAs nanowire (green), three-facet Al shell (blue), Ti/Au contacts (yellow) and bottom-gates (grey). (b) Schematic device cross section, showing orientation of applied magnetic field, $B$, and Al shell relative to the bottom-gate. (c) Magnitude of effective $g$-factor, $\left|g^*\right|$, of the lowest subgap state showing a step-like dependence on bottom-gate voltage, $V_{\rm G}$. Error bars are root-mean-square difference between upper (electron) and lower (hole) branches. (d) Differential conductance, ${\rm d}I/{\rm d}V$, as a function of source-drain bias, $V_{\rm SD}$, at gate voltage $V_{\rm g}=\SI{0.0}{\volt}$. Dashed lines correspond to $\left|g^*\right|=34$. (e)-(g) Similar to (d), but taken at gate voltage (e) $V_{\rm G}=\SI{-2.0}{\volt}$, (f) $V_{\rm G}=\SI{-4.0}{\volt}$ and (g) $V_{\rm G}=\SI{-7.5}{\volt}$, giving (e) $\left|g^*\right|=27$, (f) $\left|g^*\right|=6.6$ and (g) $\left|g^*\right|=4.3$.}
\end{figure*}

Five devices, denoted 1 to 5, were investigated. All were $\sim 2\,\mu$m long, made from MBE-grown [0001] wurtzite InAs nanowires with hexagonal cross-section \cite{Krogstrup2015}. Two devices (2 and 5) had epitaxial Al on two facets, the rest (1, 3 and 4) had epitaxial Al on three facets [Figs.~\ref{fig:1}(b), \ref{fig:2}(b) and Table~S1 in the Supplemental Material \cite{SupMaterial}]. To form a tunnel probe, the Al shell was removed by wet-etching at one end, leaving a $\sim100$~nm segment of bare InAs next to one of the normal-metal leads. The tunneling rate was controlled with the cutter-gate voltage, $V_{\rm C}$. The nanowire density in Devices 1 and 3 was controlled with bottom-gates at voltage $V_{\rm G}$ [Figs.~\ref{fig:1}(a) and S1(a) in the Supplemental Material \cite{SupMaterial}]. Device 2 used a conducting substrate at voltage $V_{\rm BG}$ [Fig.~\ref{fig:2}(a)]. Device 4 used only side-gates at voltage $V_{\rm SG}$ [Fig.~S1(b) in the Supplemental Material \cite{SupMaterial}]. Device 5 used top-gates at voltage $V_{\rm TG}$ [Fig.~S4(a) in the Supplemental Material \cite{SupMaterial}]. For all devices, gates were positioned on the side of the nanowire opposite to the Al shell. The magnetic field was oriented along the nanowire axis using a three-axis vector magnet. Standard ac lock-in techniques were used in a dilution refrigerator with a base temperature of $\sim\SI{20}{\milli\kelvin}$.

The Zeeman splitting of ABSs can be extracted from the differential conductance, ${\rm d}I/{\rm d}V$, measured as a function of applied source-drain bias, $V_{\rm SD}$, and magnetic field, $B$, along the wire. To avoid the gate-dependent level repulsion effect \cite{Lee2014}, the absolute value of the effective $g$-factor, $\left|g^*\right|$, was measured using the lowest-energy subgap state as it moved toward zero energy with $B$. Figure~\ref{fig:1}(c) shows $\left|g^*\right|$ of the lowest energy state as a function of bottom-gate voltage $V_{\rm G}$ for Device 1, displaying a characteristic step-like behavior as a function of gate voltage. A $B$-sweep at $V_{\rm G} = \SI{0.0}{V}$ displays a quasi-continuous band of ABSs with $\left|g^*\right|=34$, as shown in Fig.~\ref{fig:1}(d). The hard superconducting gap collapses at roughly $B = \SI{0.2}{\tesla}$, leaving a soft gap behind. At higher fields, the evolution of levels cannot be easily tracked. The main large gap at $V_{\rm SD}=\SI{240}{\mu\electronvolt}$---presumably arising from superconductivity among electrons that predominantly reside in the Al shell---remains visible throughout the measured range. When $V_{\rm G}$ is changed from $\SI{-2}{\volt}$ to $\SI{-4}{\volt}$, $\left|g^*\right|$ abruptly decreases from 27 to 6.6 [Figs.~\ref{fig:1}(e) and (f)]. The effective g-factor saturates at $\left|g^*\right|\sim 5$ for more negative values of $V_{\rm G}$. In contrast to the behavior at $V_{\rm G}\sim\SI{0}{V}$ where the continuum of states moved toward zero energy, evolution of single, discrete ABS can be clearly followed at $V_{\rm G} = \SI{-7.5}{V}$ [Fig.~\ref{fig:1}(e)]. In this case, the ABS with $\left|g^*\right|=4.3$ reaches zero energy at $B = \SI{1.5}{\tesla}$, with hard gaps on both sides of the state throughout the sweep.

Qualitatively similar behavior was seen in multiple device. For Device 2 at back-gate voltages in the range of $\SI{4}{\volt}$ to $\SI{6}{\volt}$, $\left|g^*\right|$ was $\sim 20$ [Fig.~\ref{fig:2}(c)]. A $B$-sweep taken at $V_{\rm BG}=\SI{4.9}{\volt}$ shows a quasi-continuous band of subgap-states with $\left|g^*\right|=19$ crossing zero-bias at $B=\SI{0.4}{\tesla}$, to become a quasi-continuum throughout the subgap region at higher field. For $V_{\rm BG}$ in the range $\SI{-2}{\volt}$ to $\SI{-8}{\volt}$, $\left|g^*\right|$ remained roughly constant at $\sim 5$. At $V_{\rm BG}=\SI{-2.4}{\volt}$, a single sharp ABS was observed, with $\left|g^*\right|=4.1$ coalescing at $B=\SI{1}{\tesla}$ and sticking to zero energy for higher fields. The narrow zero-bias conductance peak remains insensitive  to magnetic field from $1$ to $\SI{2}{\tesla}$. Line-cut plots of Figs.~\ref{fig:1}(d-g) and \ref{fig:2}(d,e) together with data from top-gated Device 5 showing a similar step-like decrease in $\left|g^*\right|$ as well as gate-voltage dependence of the effective induced superconducting gap, $\Delta^*$, are given in the Supplemental Material \cite{SupMaterial}.

We propose two contributing factors to the step-like evolution of $\left|g^*\right|$ as carriers are depleted by the gate voltage. The first is the reduction of the orbital contribution to $\left|g^*\right|$ as the wire is depleted across most of its cross section \cite{Winkler2017,Lommer1985,Kiselev1998}. The second is that the remaining density in the nanowire is predominately against the interface with the Al shell, strongly coupled to the superconductor \cite{Salis2001,Mikkelsen2018,Antipov2018}.

\begin{figure}[t]
\includegraphics[width=\linewidth]{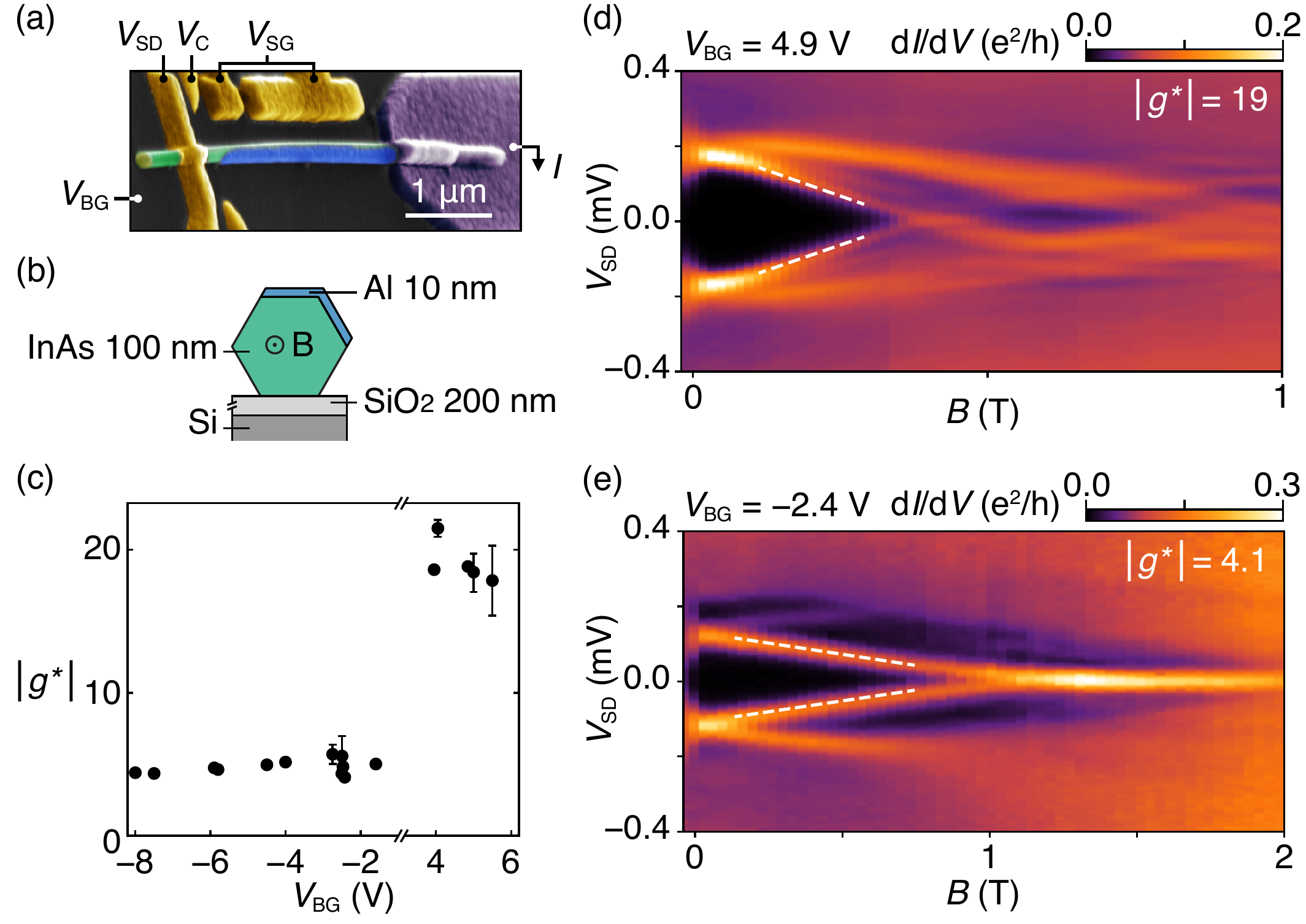}
\caption{\label{fig:2} (a) False-color electron micrograph of Device 2, consisting of InAs nanowire (green) with two-facet Al shell (blue), Ti/Au contact and side-gates (yellow), and Ti/Al/V contact (purple). (b) Schematic device cross section showing direction of applied magnetic field, $B$, and orientation of Al shell relative to the back-gate. (c) Effective $g$-factor, $\left|g^*\right|$, as a function of applied back-gate voltage, $V_{\rm BG}$. (d) Subgap state evolution in $B$, measured at $V_{\rm BG}=\SI{4.9}{\volt}$. The white, dashed lines correspond to $\left|g^*\right|=\SI{19}{}$. (e) Same as (d) taken at back-gate voltage $V_{\rm BG}=\SI{-2.4}{\volt}$, giving $\left|g^*\right|=\SI{4.1}{}$.}
\end{figure}

\begin{figure}[t]
\includegraphics[width=\linewidth]{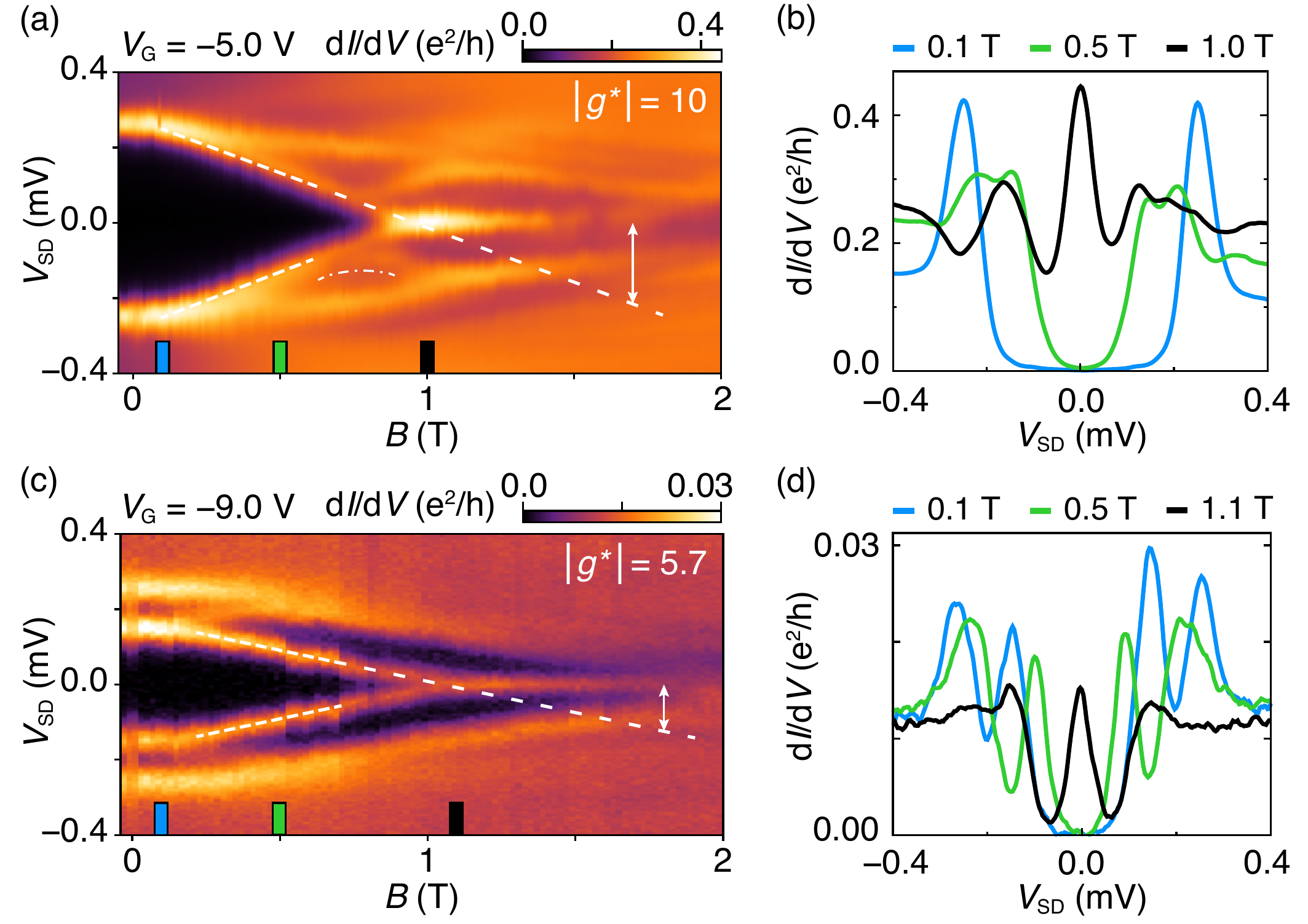}
\caption{\label{fig:3} (a) Conductance as a function of $V_{\rm SD}$ and $B$ from Device 3 taken at $V_{\rm G}=\SI{-5.0}{\volt}$. A quasi-continuous band of ABSs have $\left|g^*\right|=10$. The gap closing-reopening feature around $B=\SI{0.9}{\tesla}$ coincides with the formation of zero-bias peak persisting up to $B\sim\SI{1.7}{\tesla}$. The white arrow indicates Zeeman splitting of $\sim\SI{225}{\mu\electronvolt}$. (b) Line-cuts taken from (a) display a hard superconducting gap evolving into a zero-bias peak in high field at a subgap-state-rich regime. (c) Similar to (a) but taken at $V_{\rm G}=\SI{-9.0}{\volt}$. A discrete ABS with $\left|g^*\right|=5.7$ coalesce at zero-energy around $B=\SI{1.0}{\tesla}$. The white arrow at $B=\SI{1.8}{\tesla}$ corresponds to Zeeman splitting of $\sim\SI{125}{\mu\electronvolt}$. (d) Line-cuts taken from (c) show the emergence of a symmetric zero-bias peak with low base-conductance.}
\end{figure}

A clearer view of excited states above the lowest energy ABS, including the closing and reopening of a gap coincident with the appearance of a zero-bias conductance peak, can be seen for Device 3 in Fig.~\ref{fig:3}. Due to the gate dependent $g$-factor, it is natural to describe the robustness of the zero-bias state in the energy scale corresponding to Zeeman splitting. A $B$-sweep at $V_{\rm G}=\SI{-5.0}{\volt}$ reveals a quasi-continuous band of ABSs with a $\left|g^*\right| = \SI{10}{}$ [Fig.~\ref{fig:3}(a,b)]. At low field, the gap is hard on the low-energy side of the ABS edge, yielding small values of ${\rm d}I/{\rm d}V$; at higher fields, ${\rm d}I/{\rm d}V$ is nonzero throughout the subgap region. Around $B=\SI{0.9}{\tesla}$ an excited subgap-state (indicated by the dot-dashed line) becomes visible. It increases in energy and merges with the higher-energy ABSs around $B=\SI{1.1}{\tesla}$. The lowest energy state evolves into a zero-bias peak at roughly $B=\SI{1.0}{\tesla}$. The zero-mode can be followed up to $\sim\SI{1.7}{\tesla}$, whereafter it merges with the high subgap density. Extrapolating the $\left|g^*\right|$ slope of the lowest energy state (see the dashed line in Fig. \ref{fig:3}(a)) infers that the zero-bias peak extends for $\sim\SI{225}{\mu\electronvolt}$---comparable to the size of the main large gap. 

Lowering the gate voltage changes the picture qualitatively. The tunneling spectrum dependence on magnetic field, taken at $V_{\rm G}=\SI{-9.0}{\volt}$ displays a discrete, low-energy ABS with $\left|g^*\right|=5.7$, see Fig.~\ref{fig:3}(c). The ABS merge at $B=\SI{1.0}{\tesla}$ to form a well-defined zero-bias peak, clearly visible up to $B=\SI{1.8}{\tesla}$, corresponding to Zeeman splitting of $\sim\SI{125}{\mu\electronvolt}$. The feature of gap closing-reopening is absent in this case. The subgap conductance is low throughout the sweep [Fig.~\ref{fig:3}(d)], suggesting a low-density ABS regime.

\begin{figure}[t]
\includegraphics[width=\linewidth]{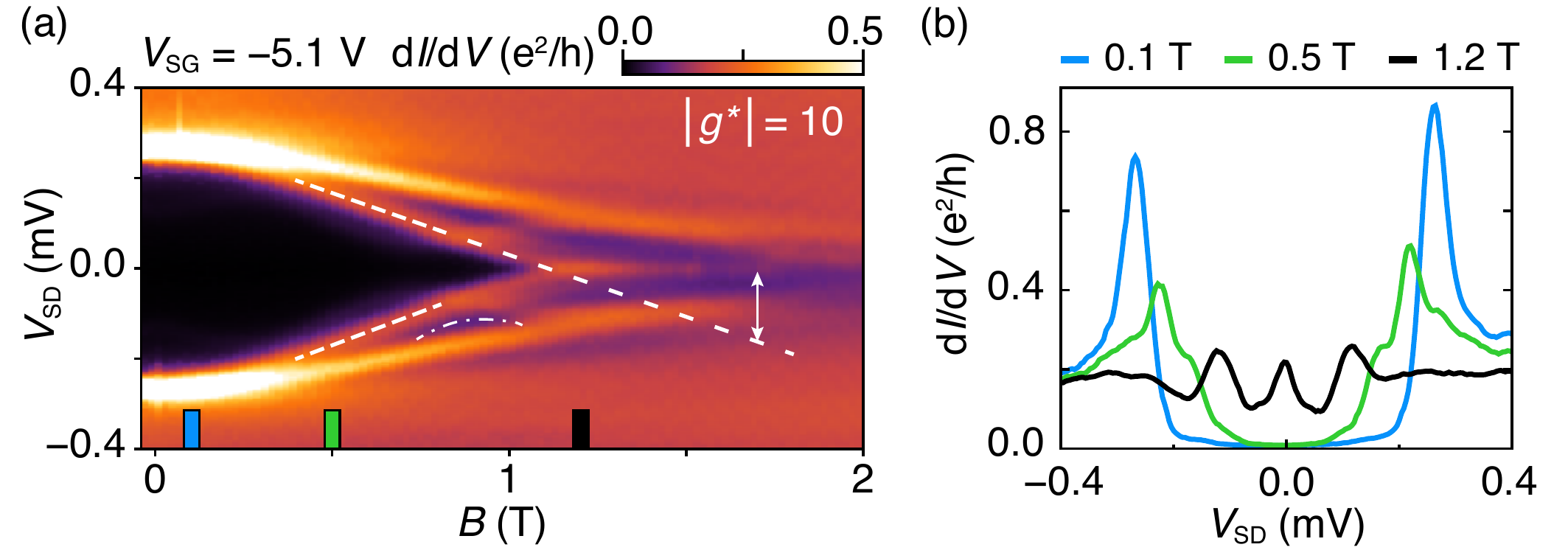}
\caption{\label{fig:4} (a) Conductance as a function of $V_{\rm SD}$ and $B$ for Device 4 at $V_{\rm SG}=\SI{-5.1}{\volt}$ shows subgap-states with $\left|g^*\right|=10$ coalescing at $B\sim\SI{1.1}{\tesla}$, while an excited state increases in energy. (b) Line-cuts taken from (a) illustrate the formation of zero-bias peak at high field. A pair of low-conductance excited states resembling gap closing-reopening are visible around $B=\SI{1.0}{\tesla}$.}
\end{figure}

The tunneling spectrum for Device 4 further illustrates the reopening of the gap [Fig.~\ref{fig:4}]. Evolution of the subgap states can be followed rather clearly: a quasi-continuous band of ABSs with $\left|g^*\right|=10$ emerges from above the gap at $B=\SI{0.3}{\tesla}$; Around $B=\SI{1.0}{\tesla}$ an excited subgap-state (indicated by the dot-dashed line) starts to gain energy with increasing field; The lowest energy state forms a zero-bias state that ranges from $B=\SI{1.1}{\tesla}$ to $\SI{1.7}{\tesla}$, corresponding to Zeeman splitting of $\sim\SI{175}{\mu\electronvolt}$ (white arrow).

The evolution of $V_{\rm SD}$ spectra with $B$ in Figs.~\ref{fig:3}(a) and \ref{fig:4}(a) show a gap to the lowest excited state that nearly closes then reopens at almost the same value of $B$ where the zero-bias peak appears. This can be interpreted as a characteristic feature of a topological phase transition \cite{Stansecu2012,Chevallier2013,Rainis2013}. The residual gap at the phase transition in both devices is finite, but less than half the energy of the main large gap---consistent with the length quantization of the wire \cite{Mishmash2016}. It has been argued theoretically \cite{Liu2017,Moore2017} and observed experimentally \cite{Lee2014,Deng2017} that a zero-bias conductance anomaly can be rendered by (partially) localized ABSs/ strongly interacting MBSs. However, numerical simulations indicate, that a topological phase transition is composed of both emergent zero-bias peak and gap closing-reopening feature \cite{Liu2017,Moore2017}. At more negative gate voltages, that is at lower electron density, the number of occupied subbands is expected to decrease. The corresponding magnetic field sweep in Fig.~\ref{fig:3}(c) shows a single ABS coalescing into a zero-bias peak, however, the gap closing-reopening feature is not visible in tunneling conductance. This is presumably due to the change in nanowire parameters, such as Rashba spin-orbit coupling, as the electric field generated by the gate voltage is increased \cite{Mishmash2016,Reeg2018}.

In summary, we have measured the effective $g$-factor of subgap states in InAs nanowires with epitaxial Al as a function of density of carriers in the wire, controlled by gate voltages, in a number of device geometries. In addition, robust zero-bias peaks---ranging for Zeeman energy comparable to the superconducting gap---have been observed at different charge carrier densities. We provide a qualitative interpretation of the data. The observations are reproduced with multiple devices. In order to understand the experimental findings in more detail, a refined electrostatic modeling considering both Zeeman and orbital contributions is desired.

We thank A.~E.~Antipov, R.~M.~Lutchyn, A.~E.~Mikkelsen and  G.~W.~Winkler for valuable discussions. This research was supported by Microsoft Research, Project Q, the Danish National Research Foundation, the Villum Foundation, and the European Commission.

S. V. and M.-T. D. contributed equally to this work.

\onecolumngrid
\clearpage
\onecolumngrid
\setcounter{figure}{0}
\setcounter{equation}{0}
\section{\large{S\MakeLowercase{upplemental} M\MakeLowercase{aterial}
}}

\renewcommand{\figurename}{FIG.~S}
\renewcommand{\tablename}{Table.~S}
\renewcommand{\thetable}{\arabic{table}}

\begin{table}[h!]
\begin{tabular*}{0.5\linewidth}{@{\extracolsep{\fill}}ccc}
\hline\hline
Device&Batch Number&Al facets\\
\hline
1 & 418 & 3\\
2 & 173 & 2\\
3 & 418 & 3\\
4 & 418 & 3\\
5 & 578 & 2\\
\hline\hline
\end{tabular*}
\caption{\label{tab:tableS1}Measured device number, corresponding nanowire growth batch number and number of Al facets covering the hexagonal InAs core.}
\end{table}

\begin{center}
\begin{figure}[h!]
\includegraphics[width=0.7\linewidth]{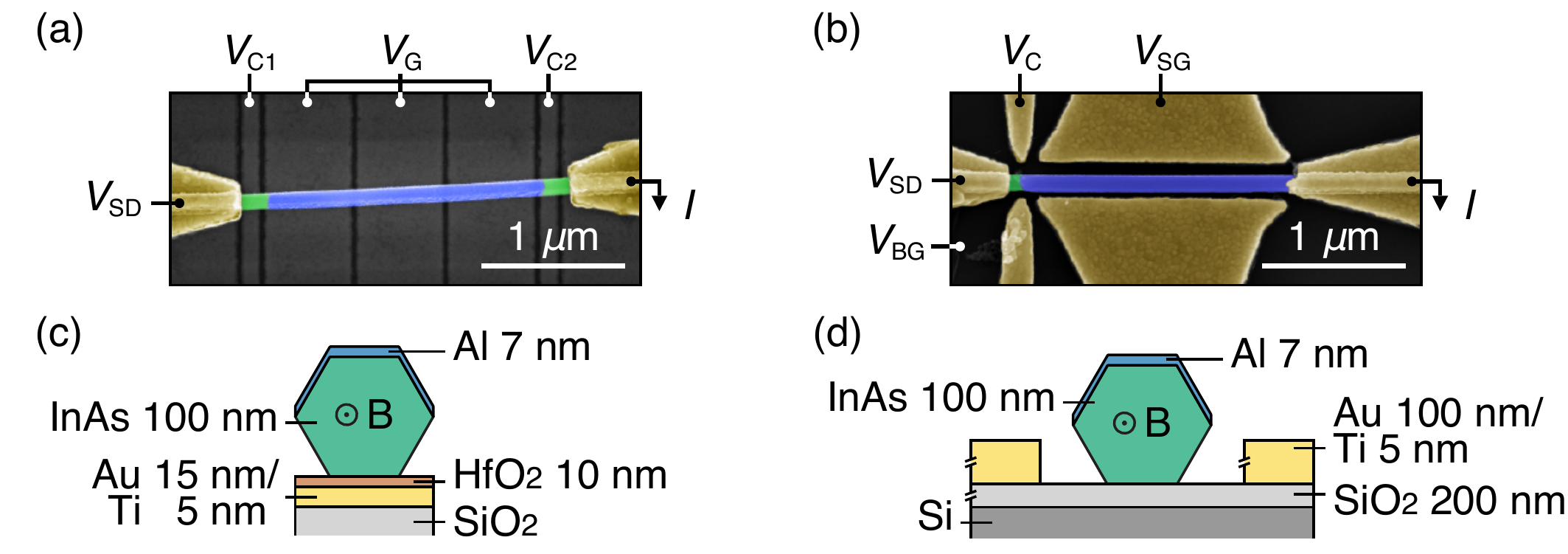}
\caption{\label{fig:S1} False-color electron micrograph of (a) device 3 and (b) device 4, showing InAs nanowire (green), three-facet Al shell (blue), Ti/Au contacts (yellow), bottom gates (grey) and side-gates (yellow). Schematic cross section of (c) device 3 and (d) device 4, showing the orientation of applied magnetic field, $B$, and Al shell relative to the gate electrodes. The tunneling spectroscopy for device 3 was performed by creating a tunneling barrier with negative voltage on $V_{C1}$, while applying $V_{C2}=\SI{1}{\volt}$, such that the other end is open to the drain. Panel (b) is adapted from Ref. \cite{supDeng2016}.}
\end{figure}
\end{center}

\begin{center}
\begin{figure}[h!]
\includegraphics[width=0.55\linewidth]{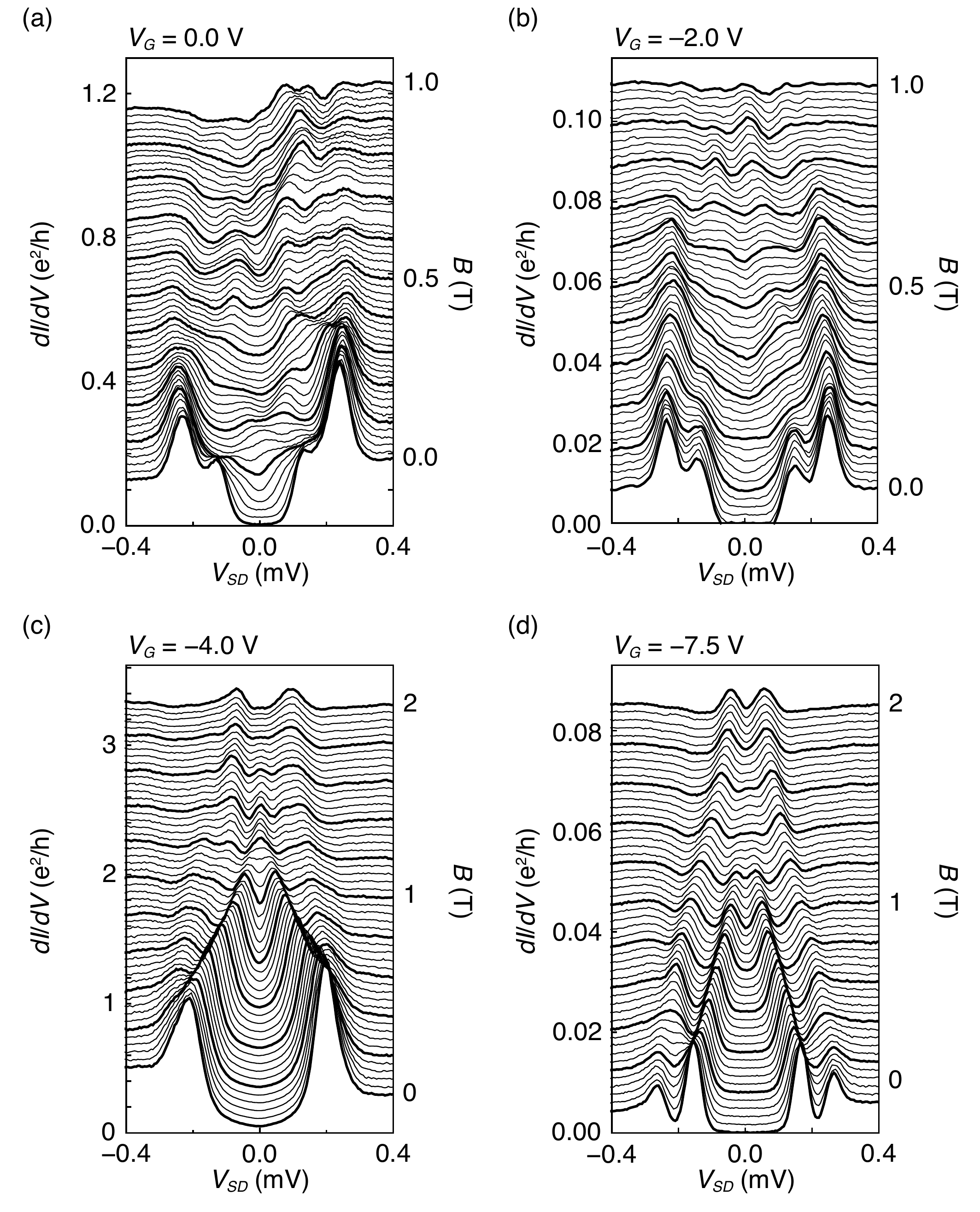}
\caption{\label{fig:S2} Waterfall plots corresponding to the data shown in the main text Fig.~1. Each curve is offset by (a)~0.02~$e^2/h$, (b) 0.004~$e^2/h$, (c) 0.06~$e^2/h$, (d) 0.016~$e^2/h$.}
\end{figure}
\end{center}

\begin{center}
\begin{figure}[h!]
\includegraphics[width=0.55\linewidth]{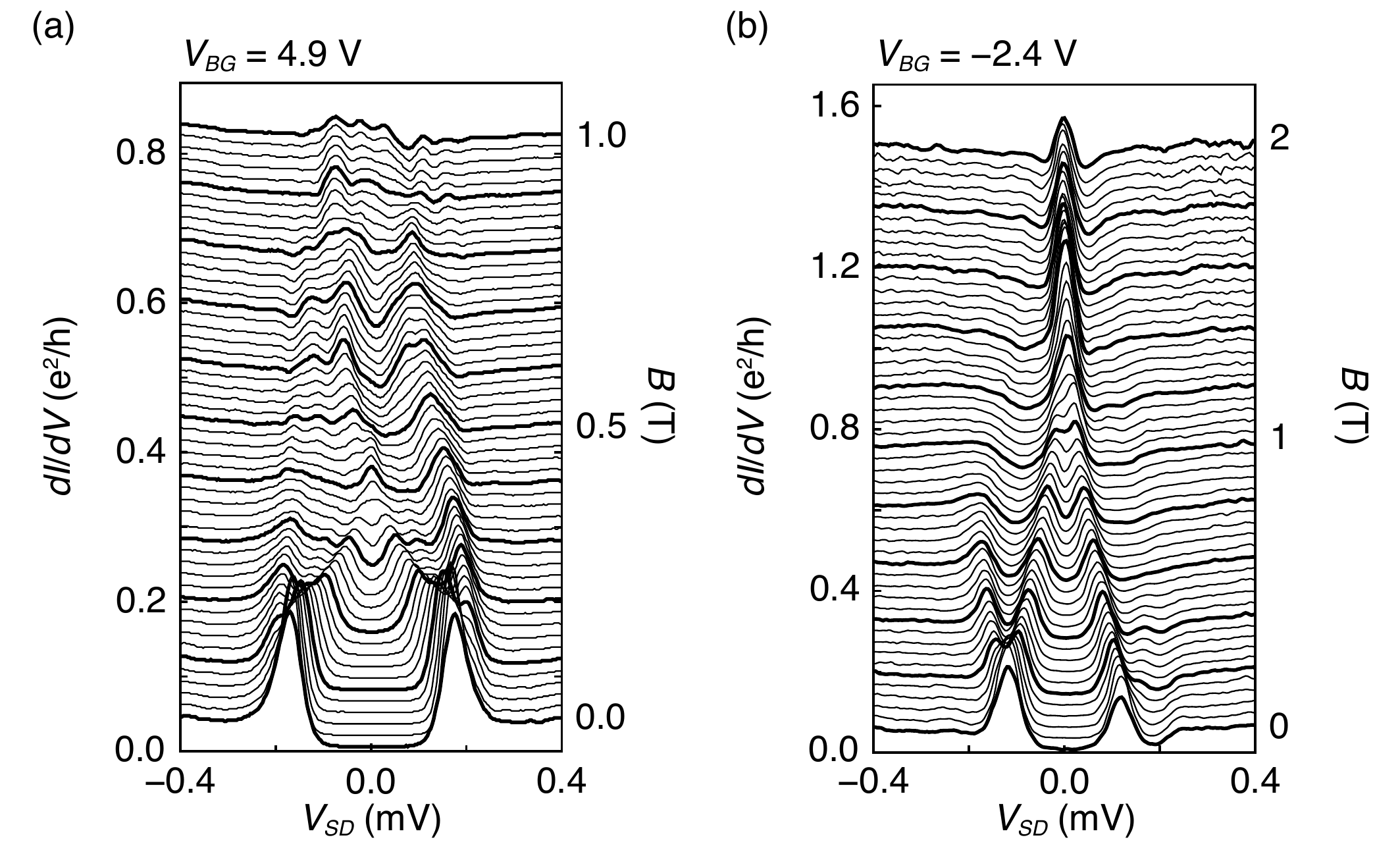}
\caption{\label{fig:S3} Waterfall plots corresponding to the data shown in the main text Fig.~2. Each curve is offset by (a) 0.015~$e^2/h$, (b) 0.0275~$e^2/h$.}
\end{figure}
\end{center}

\begin{center}
\begin{figure}[h!]
\includegraphics[width=\linewidth]{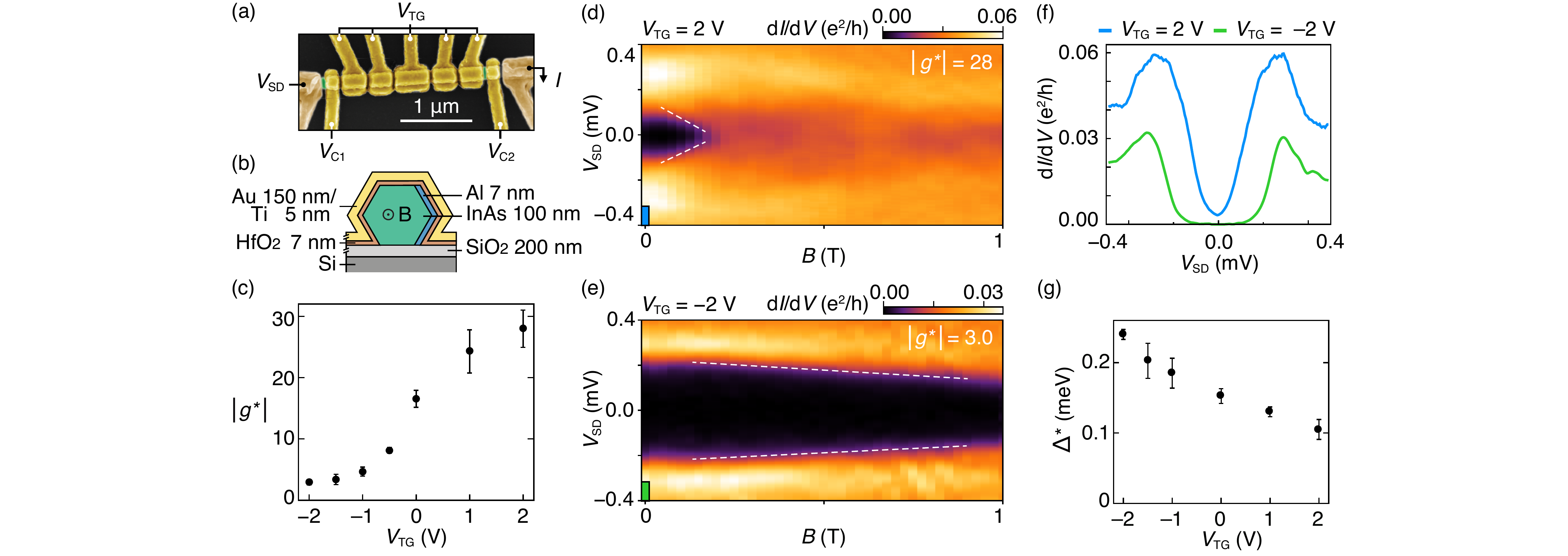}
\caption{\label{fig:S4} (a) False-color electron micrograph of device 5, showing InAs nanowire (green), Ti/Au contacts and top gates (yellow). (b) Schematic cross section of device 5, showing orientation of applied magnetic field, $B$, and Al shell relative to the top gates. (c) Magnitude of the effective $g$ factor, $\left|g^*\right|$, of the lowest subgap state showing a steplike dependence on the top-gate voltage, $V_{TG}$. (d) Differential conductance, $dI/dV$, as a function of source-drain bias, $V_{SD}$, at top gate voltage $V_{TG}=\SI{2}{\volt}$. Dashed lines correspond to $\left|g^*\right|=28$. (e) Similar to (d), but taken at top-gate voltage $V_{TG}=\SI{-2}{\volt}$, giving $\left|g^*\right|=3.0$. (f) Linecuts taken from (d) (blue) and (e) (green) at $B=\SI{0.0}{\tesla}$. (g) Effective induced superconducting gap, $\Delta^*$, as a function of $V_{TG}$, defined as the maximum slope in conductance determined by numerical derivative, $d^2I/dV^2$. Error bars in (c) and (g) are root-mean-square difference between upper (electron) and lower (hole) branches. The tunneling spectroscopy for device 5 was performed by creating a tunneling barrier with negative voltage on $V_{C1}$, while applying $V_{C2}=\SI{1}{\volt}$, such that the other end is open to the drain.}
\end{figure}
\end{center}

\newpage
\textnormal{ }

\end{document}